\pdfoutput=1

\PassOptionsToPackage{table}{xcolor}

\documentclass[10pt,journal,compsoc]{IEEEtran}
\usepackage{acro}
\usepackage{amsmath}
\usepackage{booktabs}
\usepackage{calc}
\usepackage[capitalise]{cleveref}
\usepackage{graphicx}
\usepackage{makecell}
\usepackage{multirow}
\usepackage{xfrac}
\usepackage{xspace}

\usepackage{tikz}
\usetikzlibrary{fit,positioning}

\DeclareAcronym{AoI}{short=AoI,long=application of interest,short-indefinite=an,long-indefinite=an,long-plural-form=applications of interest}
\DeclareAcronym{DTM}{short=DTM,long=dynamic thermal management}
\DeclareAcronym{DVFS}{short=DVFS,long=dynamic voltage and frequency scaling}
\DeclareAcronym{GTS}{short=GTS,long=Global Task Scheduling}
\DeclareAcronym{IL}{short=IL,long=imitation learning,short-indefinite=an}
\DeclareAcronym{IPS}{short=IPS,long=instructions per second}
\DeclareAcronym{ML}{short=ML,long=machine learning,short-indefinite=an}
\DeclareAcronym{MSE}{short=MSE,long=mean squared error}
\DeclareAcronym{NAS}{short=NAS,long=neural architecture search}
\DeclareAcronym{NN}{short=NN,long=neural network,short-indefinite=an}
\DeclareAcronym{NPU}{short=NPU,long=neural processing unit,short-indefinite=an}
\DeclareAcronym{QoS}{short=QoS,long=quality of service}
\DeclareAcronym{RL}{short=RL,long=reinforcement learning,short-indefinite=an}
\DeclareAcronym{VF}{short=V/f,long=voltage/frequency}

\DeclareMathOperator*{\argmax}{arg\,max}
\DeclareMathOperator*{\argmin}{arg\,min}

\newcommand{\ourtechacronym}{TOP-IL}
\newcommand{\ourtechplain}{\ourtechacronym\xspace}
\newcommand{\ourtech}{\emph{\ourtechacronym}\xspace}

\newcommand{\rltechacronym}{Therm-RL}

\newcommand{\rltech}{\emph{\rltechacronym}\xspace}
\newcommand{\littlec}{LITTLE\xspace}
\newcommand{\bigc}{big\xspace}
\newcommand{\gtsondemand}{\emph{GTS/\allowbreak{}ondemand}\xspace}
\newcommand{\gtspowersave}{\emph{GTS/\allowbreak{}powersave}\xspace}

\ifCLASSOPTIONcompsoc
  \usepackage[nocompress]{cite}
\else
  \usepackage{cite}
\fi
\ifCLASSINFOpdf
\else
\fi
\ifCLASSOPTIONcompsoc
 \usepackage[caption=false,font=footnotesize,labelfont=sf,textfont=sf]{subfig}
\else
 \usepackage[caption=false,font=footnotesize]{subfig}
\fi
\usepackage{url}

\begin{document}
%
\title{NPU-Accelerated Imitation Learning for\\Thermal Optimization of\\QoS-Constrained Heterogeneous Multi-Cores}
%
%
%
%

\author{Martin Rapp,
        Heba Khdr,
        Nikita Krohmer,
        and~J\"org~Henkel
\IEEEcompsocitemizethanks{%
\IEEEcompsocthanksitem M. Rapp, H. Khdr, and J. Henkel are with the Chair for Embedded Systems, Department of Computer Science, Karlsruhe Institute of Technology (KIT), 76131 Karlsruhe,
Germany.\protect\\
E-mail: martin.rapp@kit.edu, heba.khdr@kit.edu, henkel@kit.edu
\IEEEcompsocthanksitem N. Krohmer was with the Chair for Embedded Systems, Department of Computer Science, Karlsruhe Institute of Technology (KIT), 76131 Karlsruhe,
Germany. 
E-mail: nikita-krohmer@web.de%
\IEEEcompsocthanksitem A subset of this work has been first presented in DATE’22 \cite{hikey_il}.
}
}

%
%

\markboth{}%
{Rapp \MakeLowercase{\textit{et al.}}: NPU-Accelerated Imitation Learning for Thermal Optimization of QoS-Constrained Heterogeneous Multi-Cores}
%



\IEEEtitleabstractindextext{%
\begin{abstract}
\acresetall
Application migration and \ac{DVFS} are indispensable means
for fully exploiting the available potential
in thermal optimization of a heterogeneous clustered multi-core processor under user-defined \ac{QoS} targets.
However, selecting the core to execute each application and the \ac{VF} levels of each cluster
is a complex problem because 1)~the diverse characteristics and \ac{QoS} targets of applications require different optimizations, and 2)~per-cluster DVFS requires a global optimization considering all running applications.
State-of-the-art resource management techniques for power or temperature minimization either rely on measurements that are often not available (such as power) or fail to consider all the dimensions of the problem (e.g., by using simplified analytical models).
\Ac{IL} enables to use the optimality of an oracle policy, yet at low run-time overhead, by training a model from oracle demonstrations.
We are the first to employ \ac{IL} for temperature minimization under \ac{QoS} targets.
We tackle the complexity by training \iac{NN} and accelerate the \ac{NN} inference using \iac{NPU}.
While such \ac{NN} accelerators are becoming increasingly widespread on end devices, they are so far only used to accelerate user applications.
In contrast, we use an existing accelerator on a real platform to accelerate \ac{NN}-based resource management.
Our evaluation on a \emph{HiKey 970} board with an \emph{Arm big.LITTLE} CPU and \iac{NPU} shows significant temperature reductions at a negligible run-time overhead, with unseen applications and different cooling than used for training.
\end{abstract}

}

\maketitle

\IEEEdisplaynontitleabstractindextext

%
\IEEEpeerreviewmaketitle

\nocite{hikey_il}  

\IEEEraisesectionheading{\section{Introduction}\label{sec:intro}}

\IEEEPARstart{E}{levated}
on-chip temperature accelerates aging mechanisms in processors, and thereby degrades the system reliability~\cite{khdr2018aging,wang2010thermal}.
Moreover, in mobile devices, it may adversely affect the user experience since it leads to an increased skin temperature~\cite{skintemperature}.
That makes temperature minimization of paramount importance.
The two main knobs to reduce the temperature are application migration, to dynamically change the mapping of applications to cores, and \ac{DVFS}.
Using these knobs without considering the application characteristics misses significant optimization opportunities and may degrade the \ac{QoS} of the applications, thereby also degrading the user experience~\cite{pathania2018qos}.
The reason is that the impact on performance and power when migrating an application between clusters differs from one application to another~\cite{henkel2019smart}.
Similarly, the sensitivities of performance and power to \ac{DVFS} also vary.
Hence, the possibilities of \ac{QoS}-constrained thermal optimization vary between applications as the following motivational example demonstrates.

\subsection{Motivational Example}

\begin{figure}
    \centering
    \includegraphics{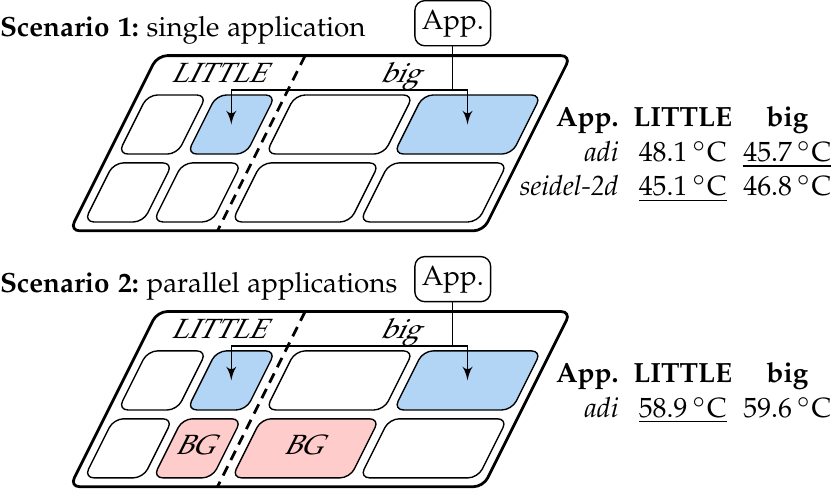}
    \caption{%
        On Arm big.LITTLE, the optimal mapping that minimizes the temperature under \acs{QoS} targets varies between applications, and with other parallel applications (BG).
        The clusters are operated at the lowest \ac{VF} levels that satisfy all \ac{QoS} targets.
    }
    \label{fig:motiv}
\end{figure}

In Scenario~1 in \cref{fig:motiv}, we execute one application, \emph{adi} or \emph{seidel-2d} from the \emph{Polybench}~\cite{polybench} suite, on an Arm big.LITTLE CPU.
The \ac{QoS} target is selected as 30\,\% of the performance, measured in \ac{IPS}, that is reached at the highest \ac{VF} level on the \bigc cluster.
The clusters are operated at the lowest \ac{VF} level that satisfies the \ac{QoS} target.
Intuitively, executing the applications on the \littlec cluster should minimize the temperature.
However, this is not always the case.
For \emph{adi}, mapping it to the \bigc cluster instead minimizes the temperature.
The reason is that \emph{adi} requires 1.8\,GHz on the \littlec cluster to reach its \ac{QoS} target, but only 0.7\,GHz on the \bigc cluster.
In contrast, \emph{seidel-2d} reaches its \ac{QoS} target already at 1.2\,GHz on the \littlec cluster, and requires 1.0\,GHz on the \bigc cluster, resulting in a similar temperature on both clusters, with a small advantage of the \littlec cluster.
The reason for the different \ac{VF} level requirements at different clusters is that the applications benefit differently from the out-of-order execution and larger caches on the \bigc cluster.
Consequently, such different application characteristics render different mappings optimal.
\emph{Optimal thermal management needs to consider application characteristics and \ac{QoS} targets.}

Scenario~2 studies \emph{adi} with the same \ac{QoS} target as in Scenario~1 but now, additional background applications with high \ac{QoS} targets run on both clusters.
Intuitively, as in Scenario~1, mapping \emph{adi} to the \bigc cluster should still minimize the temperature.
However, the background applications require to
operate both clusters at the peak \ac{VF} level to reach their \ac{QoS} targets.
Since our platform has per-cluster \ac{DVFS}, \emph{adi} is also executed at the peak \ac{VF} level.
In this case, mapping \emph{adi} to the \littlec or \bigc cluster has almost the same temperature, unlike what has been observed in Scenario~1.
Hence, per-cluster \ac{DVFS} affects the optimal mapping when several applications run in parallel.
\emph{Optimal thermal management needs to perform global optimization considering the characteristics of all running applications.}

\subsection{Challenges and Contributions}

There are several challenges in temperature minimization on heterogeneous multi-core processors under \ac{QoS} targets.
Firstly, there is high complexity in all involved aspects of the platform.
For instance, the power and performance of applications depend on the instruction sequence, CPU microarchitecture, memory architecture, and \ac{VF} level, while temperature depends on the power density, floorplan, and cooling.
Secondly, the workload, i.e., the executed applications and their arrival times, is commonly not known at design time.
Therefore, the management policy must not be specific to selected applications but achieve good management for any workload.
Thirdly, per-cluster \ac{DVFS} forces all applications on the same cluster to run at the same \ac{VF} level, requiring global optimization.
Finally, there is limited access to measurements.
For instance, most platforms, such as the one studied in this work, have no power sensors and only few temperature sensors.

Many works perform optimization with models for individual aspects such as power, performance, or temperature.
These models can be built analytically~\cite{bhat2017algorithmic} or by \ac{ML}~\cite{basireddy2019adamd,smartboost}.
However, building such models requires fine-grained access to internal measurements of processor-internal properties like power, which may not be available.
To solve this, end-to-end learning of management decisions based on the available measurements can be employed.
The two main methods to achieve this are \ac{RL} and \ac{IL}.
In both cases, \ac{NN} learning can be used to cope with the high complexity~\cite{goh1995back}.

\ac{RL} suffers from several problems.
It requires to combine objective and constraints in a single scalar reward, which does not reflect their different properties and may lead to suboptimal actions (reward hacking~\cite{reward_hacking}).
Moreover, \ac{RL} trains at run time.
This is computationally expensive, preventing a low-overhead implementation, and may result in instability such as catastrophic forgetting, leading to suboptimal management decisions.
However, run-time thermal minimization while satisfying \ac{QoS} targets requires a lightweight, yet near-optimal optimization to improve user experience, and a stable policy to avoid abrupt \ac{QoS} violations and jumps in the temperature.
\emph{\ac{IL} is the only method that provides all of these capabilities.}
In particular, it enables using the optimality of an oracle policy, which explicitly considers objectives and constraints, yet at low run-time overhead, by design-time training of a model from oracle demonstrations.
Design-time training until convergence also provides stability.
However, since \ac{IL} does not perform run-time retraining, the model must be trained such that it is capable to cope with the different scenarios that may happen at run time.
This includes for instance, different workloads, or different cooling capabilities.

Motivated by the advantages of \ac{IL}, researchers have started to apply \ac{IL} in resource management~\cite{gupta2017dypo,Kim2017Imitation,Mandal2019Dynamic,sartor2020hilite}, but they all target power or energy optimization.
This significantly differs from temperature optimization due to spatial (heat transfer) and temporal (heat capacity) effects that do not exist in power/energy.
We are the first to employ \ac{IL} for temperature optimization.

To accelerate \ac{ML}-based resource management, few works have proposed their own specific \ac{ML} accelerators~\cite{fettes2019dynamic,kwon2021reinforcement}.
However, they incur additional area overhead to the used platform and are only applicable to platforms that feature this specific accelerator.
Recently, generic \ac{NN} accelerators, e.g., \acp{NPU} or DSPs, became common in end devices such as smartphones~\cite{aibenchmark}.
These accelerators are intended to increase the performance and energy-efficiency of user applications that perform \ac{NN} inference.
Despite their increasing spread and benefits, these existing accelerators have never been used to speed up \ac{NN}-based resource management, and we are the first to do that.

We make the following novel contributions in this work:

\begin{itemize}
    \item We design, train, and employ \ac{NN}-based \ac{IL} for temperature optimization under \ac{QoS} targets, as it enables near-optimal decisions at low run-time overhead. Our solution, \ourtech, employs application migration and \ac{DVFS} on heterogeneous multi-cores.
    \item We accelerate \ourtech using an existing generic \ac{NN} accelerator (\ac{NPU}) on a real platform.
    \item We develop \ac{RL}-based thermal optimization and show that \ac{IL} outperforms \ac{RL} in terms of achieving the target objective and run-time stability.
    \item We demonstrate that the learned policy generalizes to unseen workloads and different cooling settings than what is used during training.
\end{itemize}

\section{Related Work}
\label{sec:related}

\begin{table*}
    \centering
    \caption{Overview of related work}
    \label{tab:related}
    \newcommand*{\checktikz}[1][]{\tikz[x=0.8em, y=0.8em]\draw[fill=black,draw=black,very thin,line join=round,#1] (0,.35) -- (.25,0) -- (0.8,.7) -- (.25,.1) -- cycle;}
    \newcommand{\yes}{\checktikz}
    \newcommand{\no}{$\times$}
    \newcommand{\ratheryes}[1]{(\checktikz)\textsuperscript{#1}}
    \newcommand{\easilyextendable}{\ratheryes{1}}
    \begin{tabular}{ccl|cc|cc|ccccc}
        \toprule
        \multirow{2}{*}{Technique} &
        \multirow{2}{*}{Method} &
        \multirow{2}{*}{Goal} &
        \multicolumn{2}{c|}{Actions} &
        \multicolumn{2}{c|}{Optimization} &
        Per-clust. &
        Het. &
        Unkn. &
        Multi- &
        Lim. Power \\
        &
        &
        &
        Map./Mig. &
        DVFS &
        Temp. &
        QoS &
        DVFS &
        Cores &
        Apps. &
        Prog. &
        Sensors \\
        \midrule
        \makecell{ondemand/\\powersave} &
        Rules &
        \makecell[l]{max perf./\\min P} &
        \yes &
        \yes &
        \no &
        \no &
        \yes &
        \yes &
        \yes &
        \yes &
        \yes \\
        \cite{Chen2018Profit} &
        RL &
        max perf st. P &
        \no &
        \yes &
        \no &
        \no &
        \no &
        \easilyextendable &
        \yes &
        \yes &
        \no \\
        \cite{dinakarrao2019application} &
        RL &
        min E st. R &
        \no &
        \yes &
        \no &
        \no &
        \no &
        \no &
        \yes &
        \yes &
        \no \\
        \cite{kwon2021reinforcement} &
        RL &
        min EDP &
        \no &
        \yes &
        \no &
        \no &
        \yes &
        \yes &
        \yes &
        \easilyextendable &
        \yes \\
        \cite{das2014reinforcement} &
        RL &
        max R st. QoS &
        \yes &
        \yes &
        \no &
        \yes &
        \easilyextendable &
        \easilyextendable &
        \yes &
        \no &
        \yes \\
        \cite{Donyanavard2019SOSA} &
        RL &
        min P st. QoS &
        \yes &
        \yes &
        \no &
        \yes &
        \no &
        \easilyextendable &
        \yes &
        \yes &
        \no \\
        \cite{lu2015reinforcement} &
        RL &
        min T &
        \yes &
        \no &
        \yes &
        \no &
        \easilyextendable &
        \easilyextendable &
        \yes &
        \yes &
        \yes \\
        \cite{yang2019releta} &
        RL &
        min T &
        \yes &
        \no &
        \yes &
        \no &
        \easilyextendable &
        \easilyextendable &
        \yes &
        \yes &
        \yes \\
        \cite{liu2021cartad} &
        RL &
        min T st. QoS &
        \yes &
        \yes &
        \yes &
        \yes &
        \easilyextendable &
        \easilyextendable &
        \no &
        \no &
        \yes \\
        \cite{gupta2017dypo} &
        IL &
        min E &
        \ratheryes{2} &
        \yes &
        \no &
        \no &
        \yes &
        \yes &
        \yes &
        \yes &
        \no \\
        \cite{Kim2017Imitation} &
        IL &
        min E st. QoS &
        \no &
        \yes &
        \no &
        \yes &
        \no &
        \easilyextendable &
        \no &
        \yes &
        \no \\
        \cite{Mandal2019Dynamic} &
        IL &
        min E st. QoS &
        \ratheryes{2} &
        \yes &
        \no &
        \yes &
        \yes &
        \yes &
        \yes &
        \yes &
        \no \\
        \cite{sartor2020hilite} &
        IL &
        min E st. QoS &
        \ratheryes{2} &
        \yes &
        \no &
        \yes &
        \yes &
        \yes &
        \yes &
        \yes &
        \no \\
        \midrule
        \ourtech (our) &
        IL &
        min. T st. QoS &
        \yes &
        \yes &
        \yes &
        \yes &
        \yes &
        \yes &
        \yes &
        \yes &
        \yes \\
        \bottomrule
    \end{tabular}

    \vspace{0.4ex}
    \footnotesize
    T: temperature, P: power, E: energy, R: reliability.
    \quad
    \textsuperscript{1}\,Not studied, likely applicable with minor changes.
    \quad
    \textsuperscript{2}\,Controls the number of active cores.
\end{table*}

The state-of-the-practice Android/Linux resource management~\cite{linux_gts} performs application mapping and migration (scheduling), and \ac{DVFS}.
Most schedulers are designed for homogeneous multi-core processors.
However, \ac{GTS} aims at increasing the energy efficiency of heterogeneous processors by migrating mostly-idle applications to the \littlec cluster.
Android/Linux performs \ac{DVFS} with different governors, such as \emph{powersave} for power minimization or \emph{ondemand} for a trade-off between power and performance.
However, these techniques do not consider application characteristics nor their \ac{QoS} targets, and only indirectly affect the temperature (via power or energy).

\ac{ML} provides powerful algorithms for system-level optimization~\cite{mlcad_survey}.
Supervised learning can be used to train models that predict system properties like performance or power~\cite{kim2017p4}.
Such models enable rule-based power/thermal management to predict the impact of a decision, and thereby achieve proactive management~\cite{basireddy2019adamd}.
However, model training requires access to measurements like per-core power, which are often not available in real-world processors~\cite{hikey970}.

Several works have employed \ac{RL} for power/thermal optimization~\cite{ebi2011economic}.
The works in \cite{Chen2018Profit,dinakarrao2019application,kwon2021reinforcement} use \ac{RL} for power management via \ac{DVFS}.
However, they neither consider temperature nor \ac{QoS}.
The work in~\cite{das2014reinforcement} optimizes the reliability under \ac{QoS} using both migration and \ac{DVFS}.
While reliability depends on the temperature, the two are not interchangeable.
For instance, a part of the reward function in \cite{das2014reinforcement} minimizes thermal cycling, which is unrelated to the absolute temperature.
In addition, the work does not cope with several applications running in parallel.
In~\cite{Donyanavard2019SOSA}, \ac{RL} is employed at the core level.
A high-level coordinator translates the system goal, i.e., minimizing power, into core-level target \ac{IPS}.
Then, core-level \ac{RL} agents select the \ac{VF} level to manage the core \ac{IPS} accordingly.
However, this work also does not consider temperature, is not applicable to per-cluster \ac{DVFS}, and requires run-time power measurements.
Several works employ \ac{RL} for temperature optimization.
The work in~\cite{lu2015reinforcement} performs migration for temperature minimization based on per-core temperature measurements.
In~\cite{yang2019releta}, the temperature is minimized via mapping applications at arrival time.
However, these works do not consider \ac{QoS}.
Finally, \cite{liu2021cartad} considers both temperature and \ac{QoS}.
It uses application mapping and \ac{DVFS}.
However, this work analyzes intermediate compiler-level representations of applications, and, hence, is only applicable to known applications.
In addition, it does not cope with several applications running in parallel.
\cref{tab:related} summarizes these works.

Several recent works employ \ac{IL} for system-level optimization.
The work in~\cite{gupta2017dypo} trains a model to predict the optimal number of active cores and per-cluster \ac{VF} levels to minimize the energy.
In~\cite{Kim2017Imitation}, \iac{IL} technique is proposed for \ac{DVFS} to minimize the energy under \iac{QoS} targets.
They train a separate policy per application, and, hence, cannot cope with unknown applications.
The work in~\cite{Mandal2019Dynamic} uses \ac{IL} to select the types, number, and \ac{VF} levels of active cores, for several optimization goals, e.g., minimize the energy under \iac{QoS} targets.
Finally, a hierarchical \ac{IL} technique is proposed in \cite{sartor2020hilite} to select the number of active cores and the per-cluster \ac{VF} level to maximize the energy efficiency of a heterogeneous multi-core processor under \ac{QoS} targets.
These works divide the application execution into phases and record performance counters, performance, and power for each phase at different configurations (number of active cores, \ac{VF} levels, etc.).
Oracle demonstrations are created by finding the optimal sequence of configurations per phase.
This only works because power, performance, and energy of a phase depend only on the used configuration in this phase.
However, this does not apply to temperature, which is subject to both spatial (heat transfer) and temporal (heat capacity) effects that do not exist in power/energy.
Consequently, the temperature during a phase additionally depends on all configurations of all previous phases.
This would require an exponential number of traces, which is infeasible.
In addition, the power sensors required for the oracle are often not available in real-world processors.
\cref{tab:related} also summarizes these works.
\ac{IL} has not yet been employed for thermal optimization despite its unique capabilities to combine the optimality of an oracle policy with a low run-time overhead.
We are the first to do that.

\emph{In summary, none of these works targets temperature minimization under \ac{QoS} targets, and considers heterogeneous cores with per-cluster \ac{DVFS} running parallel applications.}

\section{Problem Formulation}
\label{sec:problem}

We target a heterogeneous multi-core processor with per-cluster \ac{DVFS}, where $\mathcal{F}_x$ is the list of frequencies of cluster~$x$ and $f_x$ is its current \ac{VF} level.
There are two clusters in our platform, \littlec and \bigc, i.e., $x{\in}\{l, b\}$, but our solution is compatible with any number of clusters.
The processor executes parallel applications, each with its own \ac{QoS} target~$Q_k$ and current \ac{QoS}~$q_k$, which are expressed in terms of the \ac{IPS}.
We target an open system, where a priori unknown applications arrive at a priori unknown times.
Our solution does not rely on run-time power measurement, as they are often not available on real-world processors~\cite{hikey970}.

\begin{center}
    \begin{tabular}{ll}
        \textbf{Objective} & minimize the on-chip temperature \\
        \textbf{Constraint} & maintain \ac{QoS} of applications (\ac{IPS}) \\
        \textbf{Knobs} & app.-to-core mapping (migration), \\
        & per-cluster \ac{DVFS}. \\
    \end{tabular}
\end{center}

We split the problem into two parts: 1) application-to-core mapping (via application migration), and 2) per-cluster \ac{DVFS}.
Decisions on application migration are made with \ac{NN}-based \ac{IL}, while the \ac{DVFS} is implemented in a simple control loop.
While it would be intuitive to train a single \ac{NN} for both migration and \ac{DVFS}, performing only migration with the model reduces its complexity (create training data, topology, inference overhead).
Nevertheless, we consider \ac{VF} level information as input for migration decisions to achieve near-optimal decisions.
We accelerate the run-time inference with \iac{NPU}.
The design-time training and run-time management are described in \cref{sec:il,sec:technique:migration}, respectively.
\cref{sec:technique:dvfs} describes the \ac{DVFS} control loop.

\section{IL-based Application Migration}
\label{sec:il}

Employing \ac{IL} requires to select features, create oracle demonstrations, and train the model that is used at run time.

\subsection{Feature Selection}
\label{sec:features}

The features need to accurately describe the platform state to be able to make near-optimal migration decisions, and need to be observable at run time.
The optimal mapping of an \emph{\ac{AoI}} depends on a)~its characteristics, which affect its power and performance on different clusters, b)~its \ac{QoS} target, which determines the suitable clusters and required \ac{VF} levels, and
c)~other (background) applications, which determine the available cores, the required \ac{VF} levels per cluster to satisfy \ac{QoS} targets of the background applications, and affect the temperature distribution.

The selected features (\cref{tab:features}) cover all three aspects (a\nobreakdash-c).
The \ac{AoI} characteristics (a) comprise the current \ac{QoS} and the number of L2D accesses per second.
The latter indicates the memory-/compute-intensiveness of the \ac{AoI}.
We use the Linux \emph{perf} API to read performance counters (\ac{IPS} and L2D accesses).
The current mapping of the \ac{AoI} provides information about the source core and cluster, thereby providing context to the performance counter readings.
It is represented as one-hot encoding of all cores. 
The \ac{QoS} target (b) is represented in terms of \ac{IPS}.
The background (c) is represented by the core utilizations, as well as by the estimated \ac{VF} level change if the \ac{AoI} would not be executed (for each cluster).
The latter indicates potential temperature savings if the \ac{AoI} is migrated to another cluster.
This is calculated by first estimating the minimum \ac{VF} level~$\tilde{f}_{k,min}$ for each running application~$k$ that is required to satisfy its \ac{QoS} target~$Q_k$.
During training data generation at design time, $\tilde{f}_{k,min}$ can be determined from the execution traces.
At run time, no traces at other \ac{VF} levels are available, and  linear scaling from the current \ac{VF} level~$f_{x(k)}$ of its cluster~$x(k)$ is performed instead:
\begin{equation}
    \tilde{f}_{k,min} = \min \{ f\in \mathcal{F}_{x(k)} : q_k \cdot f / f_{x(k)} \ge Q_k \} \label{eq:f_i_min}
\end{equation}
This estimate is calculated at run time based on the current \ac{QoS}~$q_k$ in the current execution phase, i.e., \emph{$\tilde{f}_{k,min}$ does not need to be known at design time and may change over time}.
Finally, the required \ac{VF} level without the \ac{AoI} is determined per cluster~$x$ as the maximum among all other applications:
\begin{equation}
    \tilde{f}_{x\setminus AoI} = \max \{\tilde{f}_{k,min} : \text{app.~$k$ mapped to}~x \land k{\neq}AoI \}
\end{equation}

\begin{table}
    \centering
    \caption{The Selected Features for IL-based Migration (per Application)}
    \label{tab:features}
    \begin{tabular}{cc|cc}
        \toprule
        Feature & Count & Feature & Count \\
        \midrule
        AoI QoS (a)           & 1 & AoI QoS target (b)                     & 1 \\
        AoI L2D accesses (a)  & 1 & $\tilde{f}_{x\setminus AoI} / f_x$ (c) & 2 \\
        AoI curr. mapping (a) & 8 & Core utilizations (c)                  & 8 \\
        \bottomrule
    \end{tabular}
\end{table}

\subsection{Oracle Demonstrations (Training Data)}
\label{sec:training-data}

The training data need to indicate the optimal migration w.r.t.\ \ac{QoS} and temperature for a variety of scenarios.
To this end, we collect measurements of temperature and performance counters (traces) of benchmark applications in various scenarios and extract training data from the traces.

\begin{figure}
    \centering
    \includegraphics{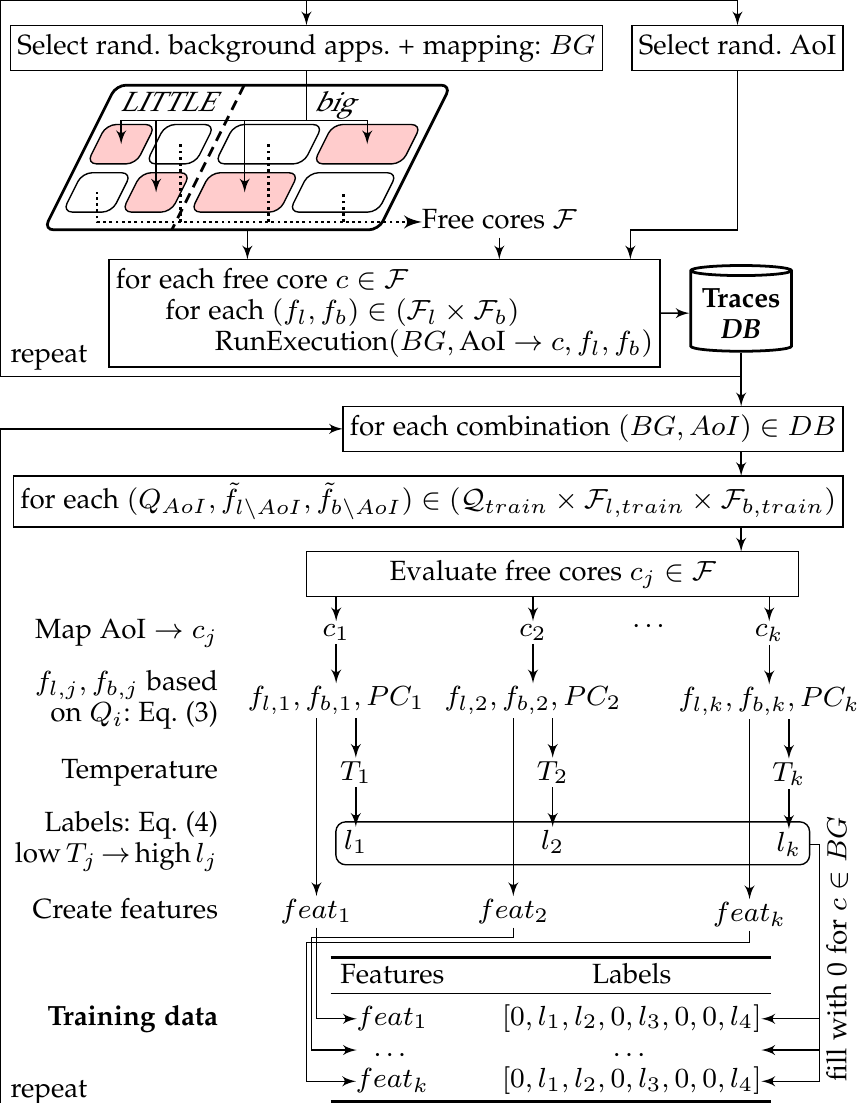}
    \vspace{-1mm}
    \caption{Design-time training data generation for \ac{IL}-based migration.}
    \label{fig:training_data}
\end{figure}

\textbf{Collect Traces:}~The process to collect traces is depicted in the upper part of \cref{fig:training_data}.
Since this is the most time-consuming part of training, redundant executions must be avoided.
The straightforward approach to collect traces would be to select a scenario, i.e., a combination of \ac{AoI}, its \ac{QoS} target, and background, and execute it once per mapping of the \ac{AoI} to each free core.
However, this creates redundant executions.
The reason is that with per-cluster \ac{DVFS}, only the application with the highest \ac{QoS} target, i.e., highest required \ac{VF} level, determines the \ac{VF} level of the cluster.
As a result, scenarios that differ only in the \ac{QoS}, may result in the same
selected \ac{VF} levels.

We avoid redundancy by obtaining traces for different combinations of per-cluster \ac{VF} levels and afterwards select different \ac{QoS} targets to create training data.
This optimization requires a constant \ac{QoS} of the benchmarks that are used to create the training data, i.e., no execution phases.
\emph{As the evaluation demonstrates, our model also generalizes to applications with execution phases.}
To further accelerate collecting traces, we stop traces after $10^{10}$ instructions of the \ac{AoI}, which is large enough to observe significant differences in the temperature between traces but still reduces the time to collect a trace, and obtain traces for a reduced set of \ac{VF} levels.
However, \ourtech supports applications with more executed instructions.
We execute the background of each scenario for 2\,min before starting the \ac{AoI} to ensure consistent initial temperature.
We randomize the order of executions to avoid any remaining systematic error.
We use active cooling with a fan because it prevents triggering \ac{DTM}, which would throttle the \ac{VF} levels unpredictably, polluting the training data.
\emph{We show in our evaluation that the trained \ac{NN} also can be used without retraining for different cooling, i.e., without a fan.}

\cref{fig:training_illu:traces:big,fig:training_illu:traces:little} present an illustrative excerpt of the collected traces (performance of the \ac{AoI} and temperature) for a single selection of background applications and \ac{AoI} (\emph{seidel-2d}).
In this example, only the two cores 3 and 6 are free.
The other cores are running background applications.

\textbf{Extract Training Data:}~The lower part of \cref{fig:training_data} shows the steps to extract training data from the collected traces: select many \ac{QoS} targets, find the corresponding traces, and create training examples.
We first select a combination of background and \ac{AoI} from the traces.
Then, we sweep the values of the \ac{QoS} target~$Q_{AoI}$ of the \ac{AoI}, and the required \ac{VF} levels of the background $\tilde{f}_{l\setminus AoI}, \tilde{f}_{b\setminus AoI}$.
Next, we
find the corresponding trace when mapping the \ac{AoI} on core~$j$ with the selected parameters.
The \ac{VF} levels~$f_l, f_b$ of this trace are the lowest
levels to satisfy~$Q_{AoI}$, $\tilde{f}_{l\setminus AoI}$, and $\tilde{f}_{b\setminus AoI}$:
\begin{equation}
    \mbox{\fontsize{9.2}{10}\selectfont\(\displaystyle %
        f_l{,}f_b{=}\argmin_{f'_l{,}f'_b} ( %
        f'_l{\ge}\tilde{f}_{l\setminus AoI}{\land} %
        f'_b{\ge}\tilde{f}_{b\setminus AoI}{\land} %
        q_{AoI}(f'_l, f'_b){\ge}Q_{AoI} \label{eq:sel_f} %
        ) %
        \)} %
\end{equation}
The peak temperature for each mapping of the \ac{AoI} to each free core~$j$ is determined from these traces.
We observe that in many cases, several mappings result in a very close temperature (e.g., mappings to different LITTLE cores).
In our experiments, there is on average one additional mapping that is within 1\,$^\circ$C of the temperature obtained with the optimal mapping.
Therefore, we use a soft label~$l_j{\in}[0,1]$, indicating the quality of mapping the \ac{AoI} to core~$j$:
\begin{equation}
    l_j = \begin{cases}
        0 & \text{core~$j$ occ.\ by background} \\
        -1 & \text{core~$j$ cannot meet~$Q_{AoI}$} \\
        e^{-\alpha (T_j - \min_{j'} T_{j'})} & \text{otherwise}
    \end{cases} \label{eq:label}
\end{equation}
Cores that are used by the background get $l_j{=}0$.
Mappings that violate the \ac{QoS} target at the highest \ac{VF} level get $l_j{=}-1$.
The mapping with the lowest temperature has~$l_j{=}1$.
For other mappings, the higher the temperature is compared to the optimum, the closer $l_j$ gets to~$0$.
The parameter~$\alpha$ determines a trade-off between tolerating slightly higher temperatures and susceptibility to temperature measurement noise.
We empirically set $\alpha{=}1$.
\cref{fig:training_illu:labels} lists some illustrative examples.
For instance, when selecting $Q_{AoI}{=}400{\cdot}10^6$\,IPS, $\tilde{f}_{l\setminus AoI}{=}1.4$\,GHz, and $\tilde{f}_{b\setminus AoI}{=}0.7$\,GHz (Line~I), the minimum frequencies of \littlec/\bigc to satisfy all \ac{QoS} targets are $1.8$\,GHz/$0.7$\,GHz and $1.4$\,GHz/$1.2$\,GHz for a mapping of the \ac{AoI} to cores 3 and 6, respectively.
This results in respective temperatures of $42.5\,^\circ$C and $46.6\,^\circ$C, i.e., a mapping to core~3 is cooler.
Therefore, the respective labels for cores~3 and~6 are $1$ and $0.02$.
\cref{fig:training_illu:labels} also lists examples where the two cores result in similar temperature, where core~6 is beneficial, and where core~3 cannot meet the \ac{QoS} target, even at the highest \ac{VF} levels (Line~II).

\begin{figure*}
    \newcommand{\named}[2]{%
        \tikz [remember picture, baseline = (#1.base)]%
        \node [inner sep=0pt, outer sep=0pt] (#1) {#2};%
    }

    \newcommand{\mips}{\,MIPS}

    \centering
    \subfloat[Trace results* (running \acs{AoI} on core 3 on the \littlec cluster)]{
        \label{fig:training_illu:traces:little}
        \noindent\begin{minipage}{0.49\linewidth}
            \centering
            \footnotesize
            \noindent\begin{tabular}{cc|cccc}
                \toprule
                \multicolumn{2}{c|}{\makebox[\widthof{Temperature~$T$}]{Performance~$q$}} & \multicolumn{3}{c}{$f_b$} & \\
                & & $0.7$\,GHz      & $1.2$\,GHz      & $1.5$\,GHz      & \ldots \\
                \midrule
                \multirow{3}{*}{$f_l$} &
                $0.5$\,GHz &
                $137$\mips &
                $140$\mips &
                $139$\mips &
                \ldots \\
                &
                $1.4$\,GHz &
                $366$\mips &
                $363$\mips &
                $373$\mips &
                \ldots \\
                &
                $1.8$\,GHz &
                \named{ex1-map3-trace-perf}{$471$\mips} &
                $478$\mips &
                $479$\mips &
                \ldots \\
                \bottomrule
            \end{tabular}\par

            \smallskip

            \noindent\begin{tabular}{cc|cccc}
                \toprule
                \multicolumn{2}{c|}{Temperature~$T$} &
                \multicolumn{3}{c}{$f_b$} & \\
                &
                &
                \makebox[\widthof{$000$\mips}]{$0.7$\,GHz} &
                \makebox[\widthof{$000$\mips}]{$1.2$\,GHz} &
                \makebox[\widthof{$000$\mips}]{$1.5$\,GHz} &
                \ldots \\
                \midrule
                \multirow{3}{*}{$f_l$} &
                $0.5$\,GHz &
                $35.8\,^\circ$C &
                $42.3\,^\circ$C &
                $50.7\,^\circ$C &
                \ldots \\
                &
                $1.4$\,GHz &
                $40.5\,^\circ$C &
                $46.2\,^\circ$C &
                $53.7\,^\circ$C &
                \ldots \\
                &
                $1.8$\,GHz &
                \named{ex1-map3-trace-temp}{$42.5\,^\circ$C} &
                $49.6\,^\circ$C &
                $56.1\,^\circ$C &
                \ldots \\
                \bottomrule
            \end{tabular}
        \end{minipage}
    }
    \subfloat[Trace results* (running \acs{AoI} on core 6 on the \bigc cluster)]{
        \label{fig:training_illu:traces:big}
        \noindent\begin{minipage}{0.49\linewidth}
            \centering
            \footnotesize
            \noindent\begin{tabular}{cc|cccc}
                \toprule
                \multicolumn{2}{c|}{\makebox[\widthof{Temperature~$T$}]{Performance~$q$}} & \multicolumn{3}{c}{$f_b$} & \\
                & & $0.7$\,GHz      & $1.2$\,GHz      & $1.5$\,GHz      & \ldots \\
                \midrule
                \multirow{3}{*}{$f_l$} &
                $0.5$\,GHz &
                $256$\mips &
                $455$\mips &
                \named{ex2-map6-trace-perf}{$563$\mips} &
                \ldots \\
                &
                $1.4$\,GHz &
                $255$\mips &
                \named{ex1-map6-trace-perf}{$455$\mips} &
                $563$\mips &
                \ldots \\
                &
                $1.8$\,GHz &
                $256$\mips &
                $454$\mips &
                $562$\mips &
                \ldots \\
                \bottomrule
            \end{tabular}

            \smallskip

            \noindent\begin{tabular}{cc|cccc}
                \toprule
                \multicolumn{2}{c|}{Temperature~$T$} &
                \multicolumn{3}{c}{$f_b$} & \\
                &
                &
                \makebox[\widthof{$000$\mips}]{$0.7$\,GHz} &
                \makebox[\widthof{$000$\mips}]{$1.2$\,GHz} &
                \makebox[\widthof{$000$\mips}]{$1.5$\,GHz} &
                \ldots \\
                \midrule
                \multirow{3}{*}{$f_l$} &
                $0.5$\,GHz &
                $38.0\,^\circ$C &
                $46.2\,^\circ$C &
                \named{ex2-map6-trace-temp}{$52.2\,^\circ$C} &
                \ldots \\
                &
                $1.4$\,GHz &
                $38.4\,^\circ$C &
                \named{ex1-map6-trace-temp}{$46.6\,^\circ$C} &
                $56.5\,^\circ$C &
                \ldots \\
                &
                $1.8$\,GHz &
                $39.5\,^\circ$C &
                $48.8\,^\circ$C &
                $57.0\,^\circ$C &
                \ldots \\
                \bottomrule
            \end{tabular}
        \end{minipage}
    }

    \subfloat[Examples for calculating the labels]{
        \label{fig:training_illu:labels}
        \centering
        \footnotesize
        \begin{tabular}{ccc|ccc|ccc|c}
            \toprule
            \multicolumn{3}{c|}{} &
            \multicolumn{3}{c|}{Trace results (\acs{AoI} on core~$3$)} &
            \multicolumn{3}{c|}{Trace results (\acs{AoI} on core~$6$)} &
            Labels \\
            $Q_{AoI}$   & $\tilde{f}_{l\setminus AoI}$   & $\tilde{f}_{b\setminus AoI}$   & $f_{l,3}$   & $f_{b,3}$   & $T_3$           & $f_{l,6}$   & $f_{b,6}$   & $T_6$           & $l_0~~\ldots~~l_7$                    \\
            \midrule
            \named{ex1-label-left}{$400$\mips}       & $1.4$\,GHz                     & $0.7$\,GHz                     & $1.8$\,GHz  & $0.7$\,GHz  & $42.5\,^\circ$C & $1.4$\,GHz  & $1.2$\,GHz  & $46.6\,^\circ$C & \named{ex1-label-right}{$0$~~$0$~~$0$~~$1.00$~~$0$~~$0$~~$0.02$~~$0$}                           \\
            $200$\mips       & $1.4$\,GHz                     & $1.2$\,GHz                     & $1.4$\,GHz  & $1.2$\,GHz  & $46.2\,^\circ$C & $1.4$\,GHz  & $1.2$\,GHz  & $46.6\,^\circ$C & $0$~~$0$~~$0$~~$1.00$~~$0$~~$0$~~$0.65$~~$0$                           \\
            $400$\mips       & $0.5$\,GHz                     & $1.5$\,GHz                     & $1.8$\,GHz  & $1.5$\,GHz  & $56.1\,^\circ$C & $0.5$\,GHz  & $1.5$\,GHz  & $52.2\,^\circ$C & $0$~~$0$~~$0$~~$0.02$~~$0$~~$0$~~$1.00$~~$0$                           \\
            \named{ex2-label-left}{$500$\mips}       & $0.5$\,GHz                     & $0.7$\,GHz                     & $-$         & $0.7$\,GHz  & $-$             & $0.5$\,GHz  & $1.5$\,GHz  & $52.2\,^\circ$C & \named{ex2-label-right}{$0$~~$0$~~$0$~~\makebox[\widthof{$0.00$}]{$-1$}~~$0$~~$0$~~$1.00$~~$0$} \\
            \bottomrule
        \end{tabular}
    }

    \subfloat[Training data examples]{
        \label{fig:training_illu:examples}
        \centering
        \footnotesize
        \begin{tabular}{cc|cccccc|c}
            \toprule            \multicolumn{2}{c|}{} & \multicolumn{6}{c|}{Features* (Excerpt)} & Labels \\
            $f_l$      & $f_b$      & $q_{AoI}$   & $Q_{AoI}$   & AoI curr. map.                  & Core utils.                     & $\tilde{f}_{l\setminus AoI} / f_l$   & $\tilde{f}_{b\setminus AoI} / f_b$   & $l_0~~\ldots~~l_7$                       \\
            \midrule
            \named{ex1-train-first-left}{$1.8$\,GHz} &
            $0.7$\,GHz &
            $471$\mips &
            $400$\mips &
            $0$~$0$~$0$~$1$~$0$~$0$~$0$~$0$ &
            $1$~$1$~$1$~$0$~$1$~$1$~$0$~$1$ &
            $0.76$ &
            $1.00$ &
            $0$~~$0$~~$0$~~$1.00$~~$0$~~$0$~~$0.02$~~$0$ \\
            $1.4$\,GHz &
            $1.2$\,GHz &
            $455$\mips &
            $400$\mips &
            $0$~$0$~$0$~$0$~$0$~$0$~$1$~$0$ &
            $1$~$1$~$1$~$0$~$1$~$1$~$0$~$1$ &
            $1.00$ &
            $0.56$ &
            \named{ex1-train-last-right}{$0$~~$0$~~$0$~~$1.00$~~$0$~~$0$~~$0.02$~~$0$} \\
            \named{ex2-train-first-left}{$1.8$\,GHz} &
            $0.7$\,GHz &
            $471$\mips &
            $500$\mips &
            $0$~$0$~$0$~$1$~$0$~$0$~$0$~$0$ &
            $1$~$1$~$1$~$0$~$1$~$1$~$0$~$1$ &
            $0.28$ &
            $1.00$ &
            $0$~~$0$~~$0$~~\makebox[\widthof{$0.00$}]{$-1$}~~$0$~~$0$~~$1.00$~~$0$ \\
            $0.5$\,GHz &
            $1.5$\,GHz &
            $563$\mips &
            $500$\mips &
            $0$~$0$~$0$~$0$~$0$~$0$~$1$~$0$ &
            $1$~$1$~$1$~$0$~$1$~$1$~$0$~$1$ &
            $1.00$ &
            $0.46$ &
            \named{ex2-train-last-right}{$0$~~$0$~~$0$~~\makebox[\widthof{$0.00$}]{$-1$}~~$0$~~$0$~~$1.00$~~$0$} \\
            \bottomrule
        \end{tabular}
    }
    \tikzset{
        frame/.style={
            draw,
            rounded corners=1.5mm,
            inner sep=2pt,
        },
        textnode/.style={
            font=\footnotesize,
            inner sep=0pt,
            outer sep=0pt,
        },
        hl1/.style={
            red,
        },
        hl2/.style={
            blue,
        },
    }
    \newcommand{\highlight}[3] {
        \node (highlight-node)
        [frame,fit=#3,#2] {};
        \node
        [textnode,#2,right=0.3mm of highlight-node] {#1};
    }
    \begin{tikzpicture}[remember picture, overlay]
        \highlight{I}{hl1}{(ex1-map3-trace-perf)}
        \highlight{I}{hl1}{(ex1-map3-trace-temp)}
        \highlight{I}{hl1}{(ex1-map6-trace-perf)}
        \highlight{I}{hl1}{(ex1-map6-trace-temp)}
        \highlight{I}{hl1}{(ex1-label-left)(ex1-label-right)}
        \highlight{I}{hl1}{(ex1-train-first-left)(ex1-train-last-right)}

        \highlight{II}{hl2}{(ex2-map6-trace-perf)}
        \highlight{II}{hl2}{(ex2-map6-trace-temp)}
        \highlight{II}{hl2}{(ex2-label-left)(ex2-label-right)}
        \highlight{II}{hl2}{(ex2-train-first-left)(ex2-train-last-right)}
    \end{tikzpicture}
    \caption{
        Illustrative example for training data generation.
        Only cores 3 and 6 are available for the \ac{AoI}.
        (a) and (b) show the trace results (\ac{AoI} performance and temperature) for the two free cores and several combinations of \ac{VF} levels $f_l$ and $f_b$.
        (c)~demonstrates the label calculation for a given \ac{AoI} \ac{QoS} target~$Q_{AoI}$, and minimum required \ac{VF} level to maintain the \ac{QoS} of the background ($\tilde{f}_{l\setminus AoI}, \tilde{f}_{b\setminus AoI}$).
        For each mapping, the minimum \ac{VF} levels that satisfy all \ac{QoS} targets are determined to obtain the temperature.
        Labels are calculated by \cref{eq:label}.
        (d)~lists some training examples.
        (I)~highlights an example, in which a mapping to the \littlec cluster is optimal.
        (II)~highlights an example, in which the \littlec can not reach the \ac{QoS} target even at the highest \ac{VF} level.
        *The number of L2 cache accesses has been omitted from the traces and features for brevity.
    }
    \label{fig:training_illu}
\end{figure*}

After creating the label, the features that describe an execution of the \ac{AoI} with the selected \ac{QoS} and background are determined from the traces according to~\cref{sec:features}.
One training example is created for each free core, where the \ac{AoI} could be executed when determining the optimal migration, i.e., each source of a migration.
This is illustrated in \cref{fig:training_illu:examples} with a few examples.
By creating one training example for \emph{every} free core for each selection of $Q_{AoI}$, $\tilde{f}_{l\setminus AoI}$, and $\tilde{f}_{b\setminus AoI}$, the process of training data generation is already exhaustive because the policy is trained to recover from each potential mapping of the \ac{AoI}.
This is the reason why we do not need to employ algorithms like \emph{DAgger}~\cite{dagger}, which initially only train the policy on the optimal sequence of management decisions, and only gradually add training data to recover from suboptimal decisions to increase the robustness of the model.
We create $19{,}831$ training examples from $100$ combinations of \ac{AoI} and background.

\subsection{IL Model Creation and Training}
\label{sec:model}

\begin{figure}
    \centering
    \includegraphics{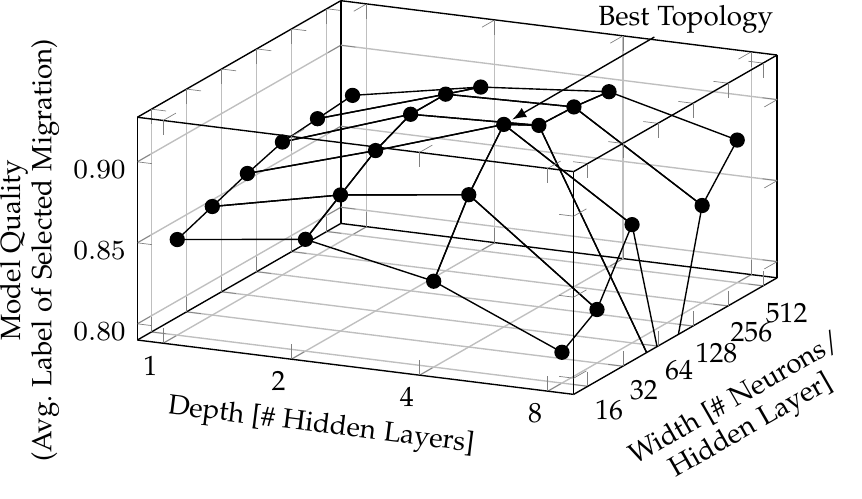}
    \caption{The best topology uses 4 hidden layers with 64 neurons, each.}
    \label{fig:gridsearch}
\end{figure}

We build a fully-connected \ac{NN} model and decide its topology (number of layers and neurons) by \ac{NAS}.
\cref{fig:gridsearch} shows the result of the grid search to determine the depth and width of the \ac{NN}.
The best topology uses 4~hidden layers with 64~neurons, each.
The hidden layers use ReLU activation, the output layer with 8 neurons does not use an activation function.
We use \emph{Adam} optimizer with momentum.
The exponentially decaying learning rate is set at $0.01{\cdot}0.95^{(epoch)}$.
We use \ac{MSE} loss and early stopping with a patience of 20~epochs.
Three models are trained with different random seed to demonstrate that the training is robust to the weight initialization, as will be shown in \cref{sec:eval}.

\section{Run-Time Temperature / QoS Management}
\label{sec:technique}

\begin{figure}
    \centering
    \includegraphics{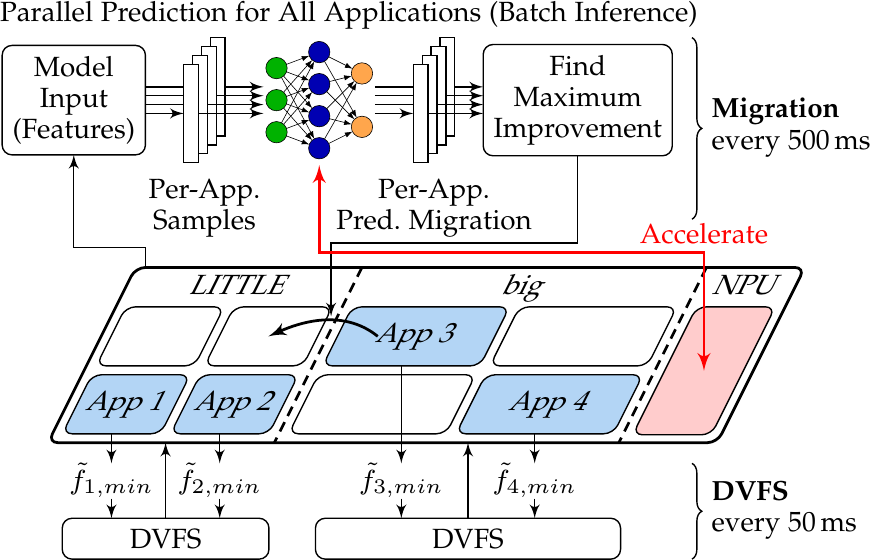}
    \caption{
        Illustration of \ourtech at run time. Application migration uses the \ac{NPU} to accelerate predicting the best migration per each application.
    }
    \label{fig:technique}
\end{figure}

The run-time part of \ourtech (\cref{fig:technique}) integrates \ac{IL}-based application migration with a per-cluster \ac{DVFS} control loop.

\subsection{Application Migration with NPU-Accelerated IL}
\label{sec:technique:migration}

If~$K$ applications run in parallel, each should be migrated to its optimal core w.r.t.\ temperature and \ac{QoS}.
However, migrating several applications at once results in a high number of potential combinations, i.e., large action space, and the impact of several migrations at once would be difficult to predict.
We solve this by migrating only one application at a time, but we find in each iteration the best migration among all possible migrations of all applications.
Our \ac{NN} model has been trained for one \ac{AoI}, which is migrated, and several other background applications.
We perform parallel inference, where each application is used as the \ac{AoI} once.
The inference output is a matrix, where each entry $\tilde{l}_{k,c}$ is the rating of mapping application~$k$ to core~$c$.
The best migration maximizes the improvement in the rating compared to the current mapping~$c(k)$:
\begin{equation}
    \hat{k},\hat{c} = \argmax_{k',c'} \left( l_{k',c'}-l_{k',c(k')} \right)
\end{equation}
The result of this optimization is to migrate application~$\hat{k}$ to core~$\hat{c}$.
The migration policy is executed each 500\,ms.
This is fast enough to adapt to changing workload phases of the applications, which run for several minutes, but still allows to maintain a reasonable overhead.

To further reduce the overhead of the \ac{NN} inference,  we employ the already existing \ac{NPU} of the \emph{HiKey 970} board.
The available parallelism in the \ac{NPU} allows performing parallel inference for all applications simultaneously in a single batch.
The \ac{NPU} is accessible via the \emph{HiAI DDK}, which originally is designed
to speed up user apps.
We develop a C++ binary that runs in user space, uses the Linux \emph{perf} API and the \texttt{/proc} filesystem to read performance counters and information about running applications, employs the \ac{NPU} for inference via the \emph{HiAI DDK} (non-blocking call), and uses the Linux \emph{affinity} feature for migration.

\begin{figure}
    \centering
    \includegraphics{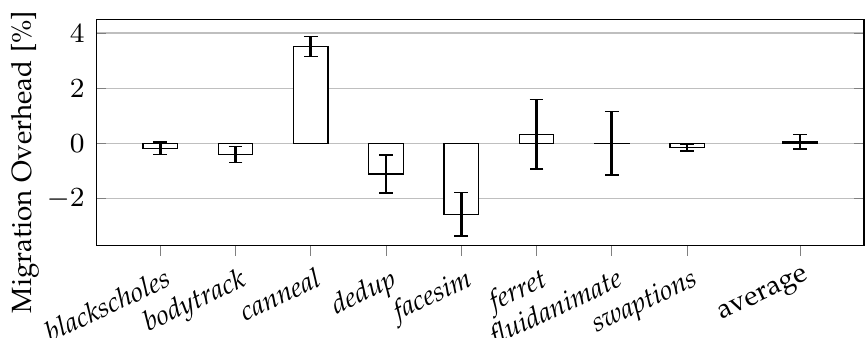}
    \caption{The performance impact of application migration is negligible.}
    \label{fig:migration_overhead}
\end{figure}

Since, we perform migration each 500\,ms, the migration overhead, e.g.,
due to cold caches, is negligible.
We perform experiments to quantify the worst-case overhead, i.e., periodically migrating an application between the \bigc and \littlec cluster in each migration epoch.
The migration overhead~$m$ is calculated by:
\begin{equation}
    m=\frac{
        \sfrac{1}{2}\cdot\left(\sfrac{1}{t_\text{big}} + \sfrac{1}{t_\text{LITTLE}}\right)  
    }{
        \sfrac{1}{t_\text{migrate}}  
    }
    - 1
\end{equation}
The numerator represents the average performance of the \bigc and \littlec clusters, while the denominator represents the measured performance with periodic migration.
We repeat each experiment three times and plot the average and standard deviation of the migration overhead of several applications in \cref{fig:migration_overhead}.
The overhead differs between applications because of their different memory and cache intensity.
For some applications, (\emph{dedup}, \emph{facesim}), we observe a negative overhead, which we interpret as follows.
If an application has different execution phases that benefit differently from the features of \bigc cluster, potential correlation between the migration epoch and the execution phases improves the performance of these applications, and thereby results in a negative overhead.
The maximum worst-case migration overhead is less than $4$\,\%, while the average worst-case migration overhead is $0.1$\,\%, which is negligible.

\subsection{Control Loop for Per-Cluster \ac{DVFS}}
\label{sec:technique:dvfs}

The \ac{IL}-based migration is integrated with \iac{DVFS} control loop to select the per-cluster \ac{VF}-levels.
The control loop utilizes the estimated~$\tilde{f}_{k,min}$ per application~$k$, as defined in \cref{eq:f_i_min}.
It then determines the minimum required \ac{VF} level per cluster~$x$ to satisfy the \ac{QoS} target of all applications running on it:
\begin{equation}
    \tilde{f}_x = \max \{\tilde{f}_{k,min} : \text{application~$k$ mapped to cluster~$x$} \}
\end{equation}
Since the run-time estimates of $\tilde{f}_{k,min}$ are based on linear scaling, they are only accurate for small \ac{VF} level changes.
Therefore, we adjust the current \ac{VF} level~$f_x$ by only one step towards~$\tilde{f}_x$ and call this control loop more frequently than migration, i.e., every 50\,ms.
We skip two iterations, one when application migration is executed and one directly after a migration,
to account for transient effects of cold caches that result in spurious \ac{QoS} violations.
Idle clusters are operated at the lowest \ac{VF} level.
We use the Linux \emph{userspace} governor to set per-cluster \ac{VF} levels.

The combination of \ac{IL}-based application migration and \ac{DVFS} control loop enables us to achieve temperature optimization under \ac{QoS} targets, as evaluated in the \cref{sec:eval}.

\section{RL-based Application Migration}
\label{sec:rl}

As discussed earlier, \ac{RL} is another method for end-to-end learning and directly making management decisions, like \ac{IL}.
However, \ac{IL} outperforms \ac{RL} in terms of stability of the learned policy.
To demonstrate this in a quantitative comparison, there is a need to implement \iac{RL}-based technique \rltech that has the same goal as our \ac{IL}-based \ourtech.
\cref{sec:related} reviewed the state-of-the-art techniques that employ \ac{RL} for application mapping/migration or \ac{DVFS}.
However, none of them targets the same goal as ours and considers heterogeneous cores with per-cluster \ac{DVFS} running parallel applications.
Therefore, this section presents \iac{RL}-based application migration policy, motivated by the state of the art, to serve as a baseline for the \ac{IL}-based policy described in \cref{sec:il}.
To enable a fair comparison between \ac{RL} and \ac{IL}, we also perform only migration with \ac{RL} and employ the same \ac{DVFS} control loop described in the previous section.

\begin{figure}
    \centering
    \includegraphics{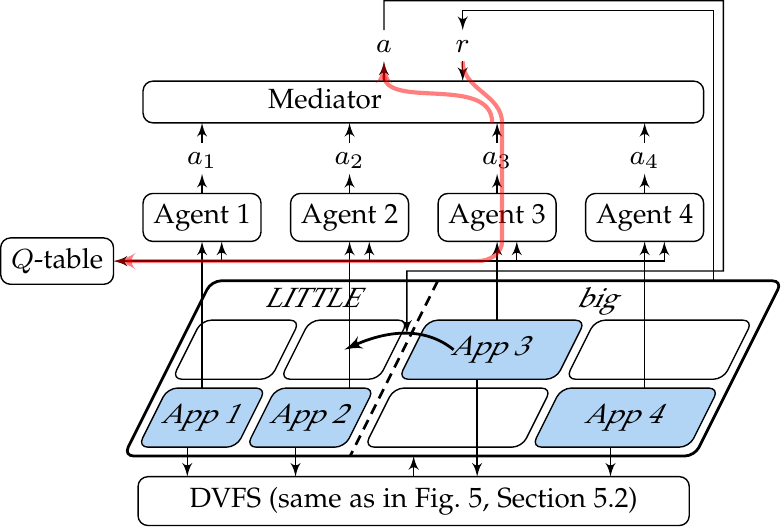}
    \caption{
        \ac{RL}-based migration instantiates one agent per application.
        All agents share the $Q$-table.
        A mediator selects the single executed action~$a$ from each per-agent action~$a_i$ (in this example from Agent~3).
        Only the selected agent updates the $Q$-table based on the reward.
    }
    \label{fig:rl_overview}
\end{figure}

\emph{TOP-IL} achieved independence from the number of running applications by performing independent inference per each running application, denoted the \ac{AoI}, to find the optimal migration.
\Ac{RL} additionally requires to perform run-time training, which requires maintaining information about the previous state.
Therefore, we instantiate one agent per application.
This has the additional benefit of maintaining state and action spaces at a reasonable size, as will be discussed in the next section.
The overall structure of \rltech is depicted in \cref{fig:rl_overview}.

\subsection{State, Action, and Reward}

The state space used for the \ac{RL} agent comprises the same features as also used for the \ac{IL} model.
In particular, these are the \ac{QoS}, number of L2D accesses, and the current mapping of the \ac{AoI}, as well as the frequencies and utilizations of the \bigc and \littlec clusters.
All these features are quantized to maintain a $Q$-table with a reasonable size.
For instance, the information about the \ac{QoS} is represented by a binary signal indicating whether or not the \ac{QoS} target is met.

The action space is selected the same as with our \ac{IL} technique, which is also the same as in \cite{lu2015reinforcement}.
There is one action per core, indicating a migration to this core, i.e., in total 8~actions.
The $Q$-table contains 2,304 entries, which is similar in size to what is reported in~\cite{kwon2021reinforcement}.

The reward function needs to combine the objective (temperature minimization) and constraint (\ac{QoS} target) into a single scalar value.
The objective is similar to~\cite{lu2015reinforcement}, which only rewards a low temperature~$T$: $r{=}80^\circ\text{C}{-}T$.
We extend it to penalize \ac{QoS} violations:
\begin{equation}
    r = \begin{cases}
        80^\circ\text{C}-T & \text{if}\ \forall i : q_i \ge Q_i \\
        -200 & \text{otherwise (\ac{QoS} violation)}\\
    \end{cases}
\end{equation}
We have empirically tuned the negative reward of $-200$ in case of \iac{QoS} violation, in order to achieve a good trade-off between low temperature and low \ac{QoS} violations.

\subsection{Multi-Agent Learning for Parallel Applications}

As discussed earlier, we instantiate one \ac{RL} agent per application.
Mediation between the agents is required to avoid 1)~contradicting decisions by different agents, and 2)~instability in the learning.
Contradicting migration decisions could result if two agents decide to perform a migration at the same time to the same core.
Such decisions should be not executed, because applications sharing a core would likely violate \ac{QoS} targets.
Moreover, even two migrations at the same time to different cores should be avoided, as simultaneous migrations might nullify the benefits of each other.
Additionally, a change in temperature when performing two migrations at once can not be traced back to either of the two, causing instability in the learning.

We, therefore, implement a mediator between the agents, similar to~\cite{jain2017cooperative}.
The mediator selects the best action among the individual actions selected by each agent based on the highest $Q$-value, and executes it.
After having executed the action, the reward obtained in the next control step should only be used to perform learning about this action, not about actions from other agents that have not been selected.
Therefore, the mediator forwards the reward only to the agent selected in the previous step to perform learning.
\cref{fig:rl_overview} illustrates the mediation process.
All agents share a common $Q$-table to improve generalization to different applications, and to immediately start with a trained policy when a new application arrives to the system.

\subsection{Training}

We select the training parameters as in~\cite{lu2015reinforcement}.
We use an $\epsilon$-greedy policy with $\epsilon{=}0.1$, a discount factor $\gamma{=}0.8$, and a learning rate~$\alpha{=}0.05$.
As the $Q$-table is initialized with constant values, a high-quality \ac{RL} policy is only obtained after significant training.
Therefore, the initial performance of \iac{RL} policy is not representative.
We avoid this by first training a policy until convergence ($\sim$3\,h) on a different random workload from what is used later in the evaluation.
We then store the $Q$-table and load it at the beginning of each evaluation run.
To reduce the impact of randomness on the policy performance, three policies are trained with different random seeds, like with the \ac{IL} model.

\section{Experimental Evaluation}
\label{sec:eval}

We perform experiments on a \emph{HiKey970}~\cite{hikey970} board.
It employs a HiSilicon Kirin~970 smartphone SoC that implements the common Arm big.LITTLE architecture with four Arm Cortex-A53 and four Arm Cortex-A73 cores.
It supports per-cluster \ac{DVFS} with frequencies up to 1.84\,GHz and 2.36\,GHz, respectively.
Furthermore, it comes with \iac{NPU} to accelerate \ac{NN} inference.
The board runs Android~8.0.
We place the board in an A/C room to maintain a constant ambient temperature.
The on-chip temperature is monitored with the on-board thermal sensor with a frequency of 20\,Hz.

\ourtech is compared with \rltech presented in \cref{sec:rl}, as well as with state-of-the-practice solutions, Linux \ac{GTS}, paired with either \emph{ondemand} or \emph{powersave} governors.
\ac{GTS} assigns applications to a cluster depending on the computational requirements, i.e., mostly-idle and performance-hungry applications are migrated to the \littlec and \bigc cluster, respectively.
\emph{Ondemand} aims at providing a high performance but saving power when low performance is required.
It achieves this by scaling the \ac{VF}-levels according to the CPU utilization, where \ac{VF} levels are upscaled if the utilization exceeds a fixed threshold, and downscaled if it falls below a second threshold.
\emph{Powersave} minimizes the power consumption by always operating at the lowest \ac{VF} levels, irrespective of the associated performance losses.
These Linux policies are not aware of detailed application characteristics or \ac{QoS} targets.
\gtsondemand is the default configuration that is shipped with Android~8.0 on \emph{HiKey970}.

\textbf{Generalization and Robustness:~}
We demonstrate that \ourtech and the employed \ac{NN} model can cope with:
1)~\emph{Unseen applications} that have not been used for training.
2)~\emph{Different cooling}: We perform experiments also with passive cooling (without a fan) instead of the active cooling used for training data generation.
3)~\emph{Randomness in the training and at run time}:
We train three models with different random seeds to demonstrate the robustness to weight initialization.
We then repeat the experiments three times, where each repetition uses a different model, and report average and standard deviation of results.
This demonstrates robustness to run-time variability due to
workload fluctuations.
In addition, we demonstrate 4)~the \emph{stability} of the learned policy.

\subsection{Illustrative Example}

\begin{figure*}
    \centering
    \subfloat[\emph{adi} (*optimal mapping: \bigc)]{
        \label{fig:motiv_illu:adi}
        \includegraphics{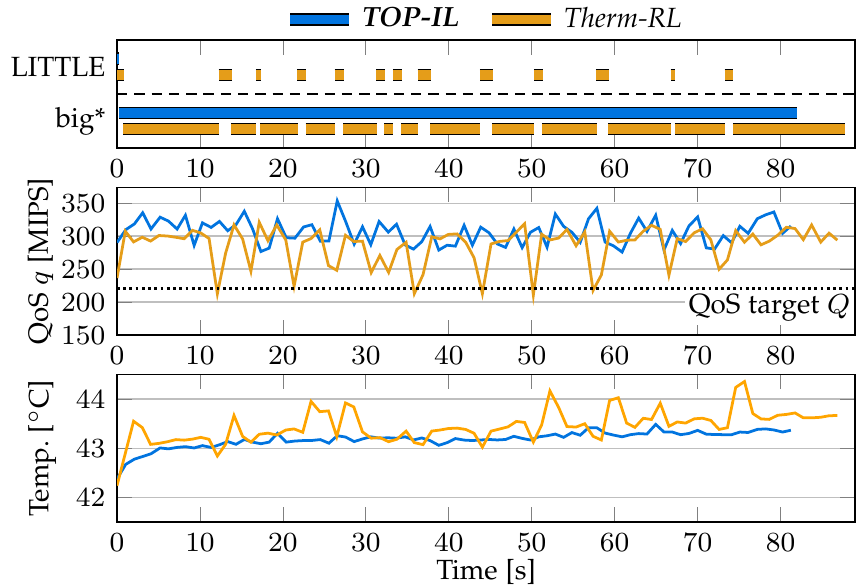}%
    }
    \subfloat[\emph{seidel-2d} (*optimal mapping: \littlec)]{
        \label{fig:motiv_illu:seidel}
        \includegraphics{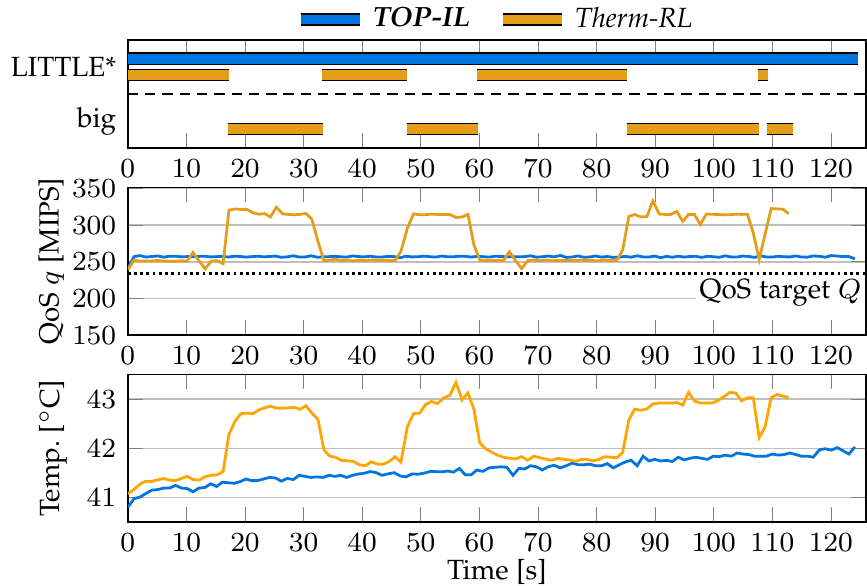}%
    }
    \caption{
        Illustrative example demonstrating the mappings chosen by our \ourtech and \rltech with the two applications \emph{adi} and \emph{seidel} studied already in \cref{fig:motiv}.
        Our \ourtech selects the optimal mapping for both applications.
        \rltech in general shows a similar trend but is unstable, selecting also suboptimal mappings.
        The \ac{QoS} targets are reached in all cases.
        However, \rltech increases the temperature during suboptimal mappings.
    }
    \label{fig:motiv_illu}
\end{figure*}

\begin{figure*}
    \centering
    \subfloat[With a fan (same as for oracle demonstrations)]{
        \label{fig:sota:fan}
        \includegraphics{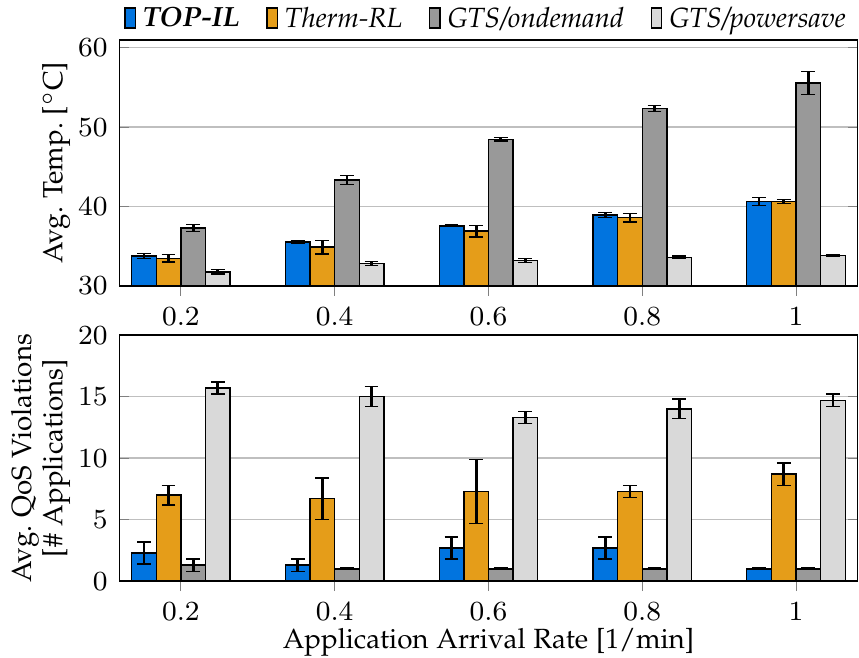}%
    }
    \subfloat[Without a fan]{
        \label{fig:sota:nofan}
        \includegraphics{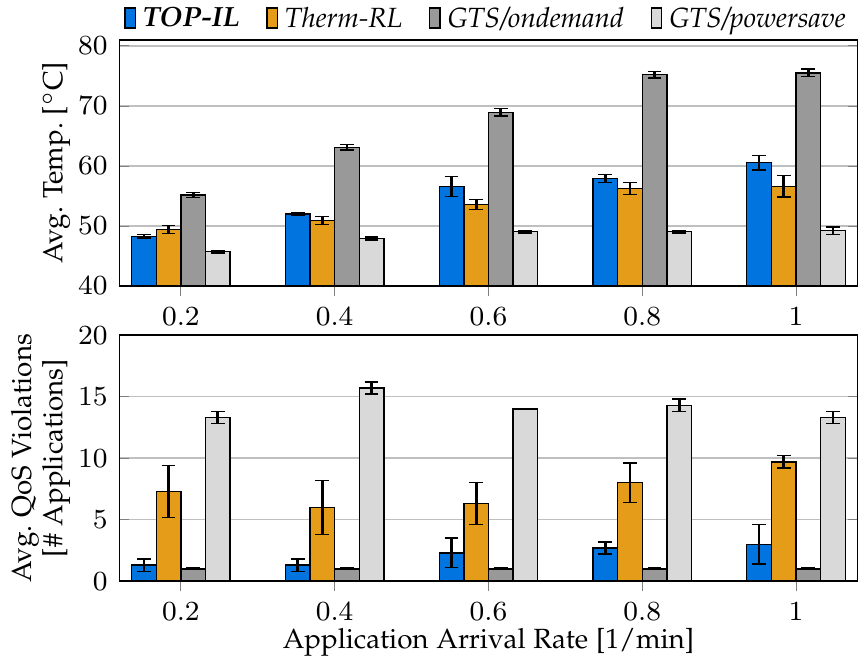}%
    }
    \caption{
        Main results: Our \ourtech significantly reduces the temperature, while achieving low QoS violations.
        This is the case both when running with a fan, as when recording the traces for the oracle demonstrations, but also without the fan, demonstrating the generalization of our model.
        Bars show mean and standard deviation over three experiments.
        \ourtech and RL use models trained with different random seeds.
    }
    \label{fig:sota}
\end{figure*}

We first present an illustrative example comparing the migration decisions of \ac{IL} and \ac{RL}.
We study the same case as presented in the motivational example in \cref{fig:motiv}, i.e., we run the two applications \emph{adi} and \mbox{\emph{seidel-2d}}.
\cref{fig:motiv_illu:adi} shows the selected cluster (mapping) of \emph{adi}.
A mapping to the \bigc cluster is optimal. 
\ourtech always selects the optimal mapping.
\rltech also mostly selects a mapping to the \bigc cluster but infrequently migrates \emph{adi} to the \littlec cluster.
In both cases, \emph{adi} reaches its \ac{QoS} target.
The temperature reached by the two techniques is also similar, as they select the same mapping most of the time.
\cref{fig:motiv_illu:seidel} shows the mappings selected with \emph{seidel-2d}, for which the \littlec cluster is optimal.
\ourtech again consistently selects the optimal mapping.
In contrast, \rltech is more unstable and migrates \mbox{\emph{seidel-2d}} irregularly between both clusters.
This results in an unnecessarily high \ac{QoS} during the time on the \bigc cluster, which also results in a higher temperature during these periods.
These examples illustrate that the policy learned with \ac{IL} is stable and consistently selects the optimal mapping, in contrast to \ac{RL}, which is more unstable.
This ultimately results in a lower temperature.
The instability of \ac{RL} leads to even worse results (\ac{QoS} violations) with more realistic workloads with multiple parallel applications, as will be shown in the next section.

\subsection{Main Experiment: Parallel Mixed Workload}
\label{sec:eval:sota}

We now evaluate the capabilities of all techniques to reduce the temperature under \ac{QoS} targets.
We create a mixed workload of 20~randomly selected applications among  \emph{blackscholes}, \emph{bodytrack}, \emph{canneal}, \emph{dedup}, \emph{facesim}, \emph{ferret}, \emph{fluidanimate}, and \emph{swaptions} from PARSEC~\cite{parsec}, and \emph{adi}, \emph{fdtd-2d}, \emph{floyd-warshall}, \emph{gramschmidt}, \emph{heat-3d}, \emph{jacobi-2d}, \emph{seidel-2d}, and \emph{syr2k} from \emph{Polybench}~\cite{polybench}.
Only the \emph{Polybench} applications (except \emph{jacobi-2d}) have been used for training \ourtech\ and \rltech.
All other applications are unseen.
We select a random \ac{QoS} target for each application.
The arrival times are distributed by a \emph{Poisson} distribution with varying arrival rate to test different system loads.
With \ourtech, the average/peak system utilizations vary from 13\,\%/38\,\% to 37\,\%/75\,\%, for minimum and maximum arrival rates, respectively.
We let the board cool down for 10\,min between experiments.
All experiments are performed three times (with different models for \ourtech and \rltech), as explained earlier.

\cref{fig:sota:fan,fig:sota:nofan} shows the results (mean and standard deviation for three repetitions) for the cooling with a fan, i.e., like for training data generation, \emph{and without a fan}, i.e., different from the training data, respectively.
\ourtech reduces the average temperature by up to 17\,$^\circ$C compared to \gtsondemand at only slightly more \ac{QoS} violations.
\gtspowersave achieves the lowest temperature but the majority of applications violate their \ac{QoS} target.
Finally, the temperature with \rltech is similar to \ourtech.
However, \ourtech achieves 63\,\% to 89\,\% fewer \ac{QoS} violations.
\ourtech is the only technique to achieve temperature minimization at few \ac{QoS} violations.
This result is independent of the cooling.

\begin{figure}
    \centering
    \includegraphics{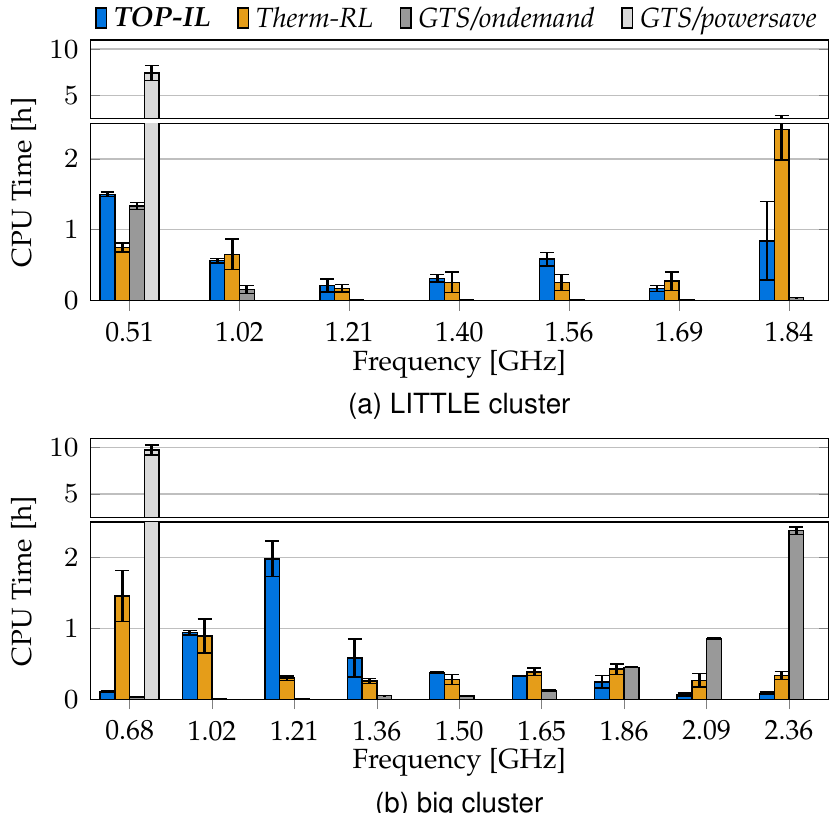}
    \caption{Total CPU time (among all arrival rates) per cluster and \ac{VF} level per technique in the experiments without a fan (\cref{fig:sota:nofan}).}
    \label{fig:frequency_usage}
\end{figure}

To explain these results we analyze the selected mappings and \ac{VF} levels.
\cref{fig:frequency_usage} plots the distribution (mean and standard deviation for the three repetitions) of the total CPU time (time executing an application) for executing the workload at all arrival rates according to the cluster and selected \ac{VF} level for the experiment without a fan.
\ac{GTS} favors the \bigc cluster and \emph{ondemand} selects high frequencies when applications are executed.
As a result, \gtsondemand uses most CPU time at the highest \ac{VF} level on the \bigc cluster, leading to low \ac{QoS} violations.
However, this also leads to high temperature and ultimately even causes thermal throttling, forcing \gtsondemand to occasionally reduce the \ac{VF} levels.
In contrast, \emph{powersave} always selects the lowest \ac{VF} level.
The reduced performance increases the number of simultaneously running applications, which forces \ac{GTS} to also use the \littlec cluster.
As a result, \gtspowersave uses CPU time on both clusters at the lowest \ac{VF} level, leading to the lowest temperature but many \ac{QoS} violations.
\rltech uses a lot of CPU time on the \littlec cluster at the highest \ac{VF} level and on the \bigc cluster at the lowest \ac{VF} level.
In both cases, a migration to the other cluster would likely have been beneficial to either be able to satisfy the \ac{QoS} target, or to reduce the temperature.
In particular, the high CPU time spent on the \littlec cluster at peak \ac{VF} level explains the high number of \ac{QoS} violations.
The reason for the suboptimal mapping decisions of \rltech are policy instability due to continual exploration in online learning and combining objectives and constraints into a single scalar reward.
In contrast, \ourtech uses more time on the \bigc cluster at rather low \ac{VF} levels, which allows it to meet the \ac{QoS} target at a low temperature, as seen in \cref{fig:sota}.
We also did this analysis for the experiment with a fan and found similar results (except for no throttling with \gtsondemand).
\emph{In summary, \ourtechplain is the only technique to achieve temperature minimization at low \ac{QoS} violations.
This is achieved for mixed workloads containing unseen applications, for different cooling setting than used during training, and is reproducible for models trained with different random initialization.}

\subsection{Single-Application Workloads}

\begin{figure}
    \centering
    \includegraphics{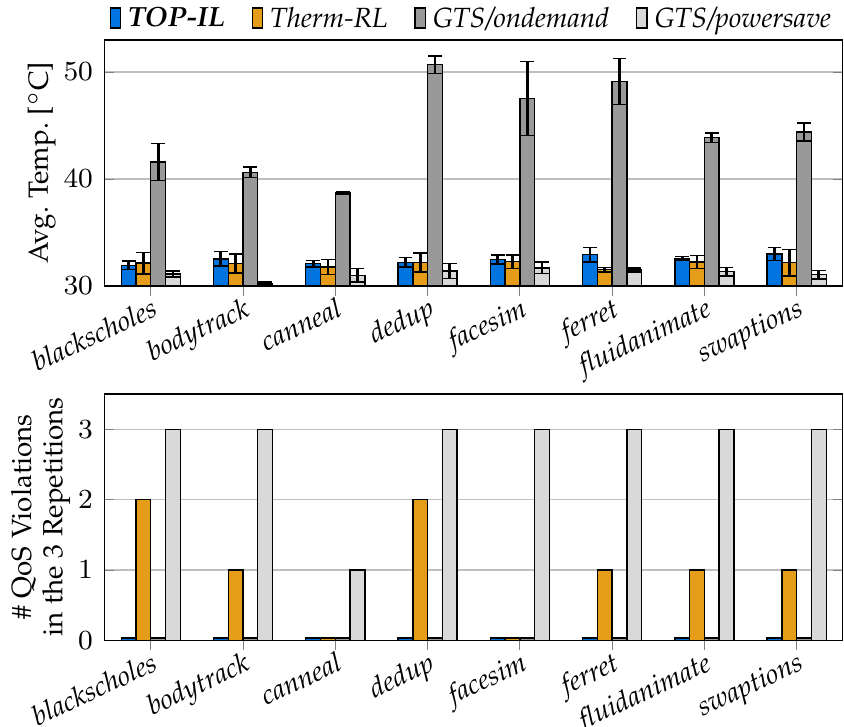}%
    \caption{
        Our \ourtech is the only technique to achieve no performance violations, yet low temperature for all single application workloads.
        All applications are unseen, i.e., not used for training.
    }
    \label{fig:sota_single}
\end{figure}

The results of the previous section contain both seen and unseen applications.
To further demonstrate the generalization, we run experiments with \emph{only unseen} applications.
The \ac{QoS} targets are set such that they can be met at the highest \ac{VF} level on the \littlec cluster.
As in the previous section, we repeat each experiment three times with different \ac{IL} or \ac{RL} models.
\cref{fig:sota_single} visualizes the results in terms of average temperature and \ac{QoS} violations.
As in the previous experiments, \gtsondemand reaches the highest temperature.
The other three techniques all result in a similar low temperature.
As there is only one application per workload, it can either reach or violate its \ac{QoS} target.
We therefore report the number of executions with \iac{QoS} violation instead of the average number of applications that violate their \ac{QoS}.
As expected, \gtspowersave violates almost all \ac{QoS} targets.
The only exception is \emph{canneal}, which is memory-intensive and its performance depends less on the CPU \ac{VF} level.
\rltech also violates the performance constraint in 33\,\% of the executions.
The reason is that the policy learned with \ac{RL} suffers from instabilities, which causes frequent migrations.
After each migration, the \ac{DVFS} control loop requires a few iterations to determine the \ac{VF} level.
During this time, the \ac{QoS} may be temporarily violated, potentially resulting in a global \ac{QoS} violation among the whole execution.
The only technique that achieves both a low temperature and no \ac{QoS} violations is \ourtech.
These experiments demonstrate again the capabilities of \ourtech to effectively minimize the temperature under \iac{QoS} target, but most importantly also \emph{the generalization capabilities of \ourtechplain to unseen applications}.

\subsection{Model Evaluation}

This section evaluates the \ac{NN} model in isolation.
We split the training/test data into training and test based on the \ac{AoI}, where seven out of nine benchmarks are only used for training (same as in previous sections), and others only for testing.
As discussed earlier, our goal is to select any near-optimal mapping in case several mappings result in a similar temperature.
The following reports the mean and standard deviation across three models trained with different random seeds.
Our model selects a mapping within 1\,$^\circ$C of the optimum in $82{\pm}5\,\%$ of the cases.
The selected mapping is, on average, only $0.5{\pm}0.2\,^\circ\text{C}$ hotter than the optimum.
\emph{This demonstrates that our training process is robust and consistently creates models that make near-optimal decisions.}

\subsection{Run-Time Overhead}

\begin{figure}
    \centering
    \includegraphics{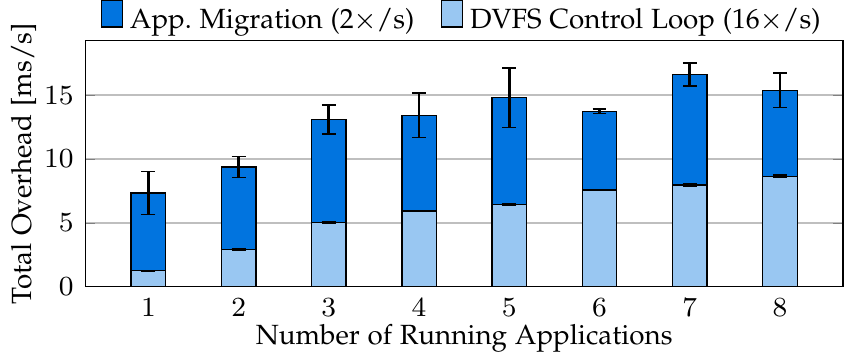}%
    \caption{
        The overhead of the \acs{DVFS} control loop increases with the number of executed executions, whereas
        application migration has a constantly low overhead due to the parallel \acs{NN} inference with the \acs{NPU}.
    }
    \label{fig:overhead}
\end{figure}

The results in \cref{fig:sota,fig:frequency_usage,fig:sota_single,fig:motiv_illu} already inherently contain the run-time overhead (additional CPU load, induced temperature) of \ourtech as it is running in parallel to the workload.
We perform in this section additional experiments to explicitly evaluate the overhead of our technique.
We study different system utilization values, i.e., different numbers of running applications.
\cref{fig:overhead} presents the results.
The \ac{DVFS} control loop is executed 16 times per second.
Its overhead increases with the number of managed applications.
The main component is reading the performance counters, which scales linearly with the number of applications.
In contrast, the overhead of the migration policy, which is executed twice per second, barely changes with more running applications.
This is as its main component is the \ac{NN} inference, which uses parallel inference of the \ac{NN}, and thereby maintains a constant low latency.
In the worst case, the \ac{DVFS} control loop and migration policy have an overhead of 8.7\,ms/s and 8.6\,ms/s (0.54\,ms and 4.3\,ms per invocation), respectively.
The total run-time overhead of \ourtech is $\le$1.7\,\%, and therefore negligible.
It is important to notice that we use a single-threaded implementation of \ourtech, i.e., the overhead only affects a single core.

\section{Conclusion}

Temperature minimization under \ac{QoS} targets requires application migration and \ac{DVFS}.
Optimization can only be achieved by jointly considering the diverse characteristics and \ac{QoS} targets of all running applications, and, hence, is a complex problem.
We tackle the complexity with \ac{NN}-based \ac{IL}, which enables us to combine the optimality of the oracle policy with a low run-time overhead.
We employ the existing \ac{NPU} of a smartphone SoC to accelerate the run-time inference.
Our policy offers stable management and generalizes to different workloads and cooling settings than what has been used for training.

\ifCLASSOPTIONcompsoc
  \section*{Acknowledgments}
\else
  \section*{Acknowledgment}
\fi
This work was partly funded by the Deutsche Forschungsgemeinschaft (DFG, German Research Foundation) -- Project Number 146371743 -- TRR 89 Invasive Computing.

\ifCLASSOPTIONcaptionsoff
  \newpage
\fi


\bibliographystyle{IEEEtran}
\bibliography{bibliography}

%

\begin{IEEEbiography}[{\includegraphics[width=1in,height=1.25in,clip,keepaspectratio]{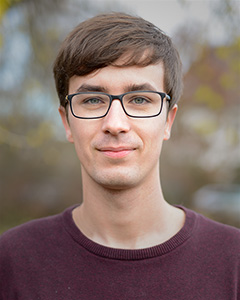}}]{Martin Rapp}
has successfully defended his Ph.D. in Computer Science at Karlsruhe Institute of Technology (KIT) in May 2022 under the supervision of Prof.\ Dr.\ J\"org Henkel.
Mr.~Rapp received a B.Sc. and M.Sc. degree -- both with distinction -- in Computer Science from the KIT in 2014 and 2016, respectively.
His current research focuses on resource-constrained machine learning: ML-based run-time resource management for many-core architectures and distributed resource-aware on-device training of neural networks.
ORCID 0000-0002-5989-2950
\end{IEEEbiography}

\begin{IEEEbiography}[{\includegraphics[width=1in,height=1.25in,clip,keepaspectratio]{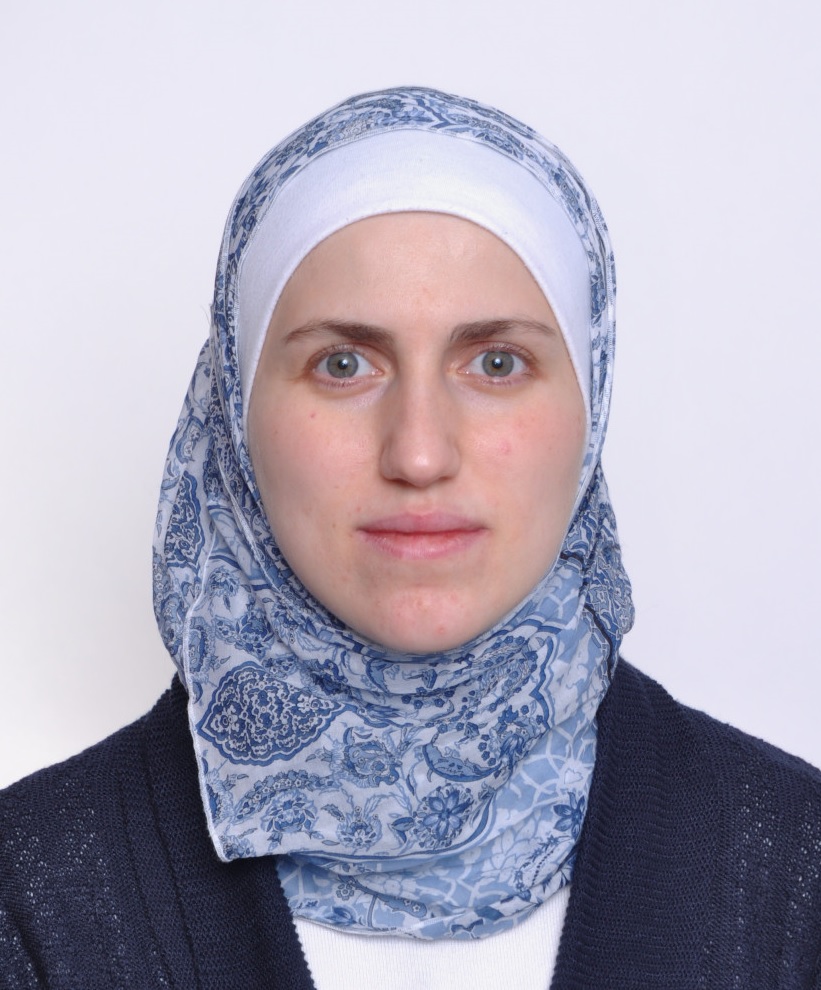}}]{Heba Khdr}
received her Ph.D. (Dr.-Ing.) in Computer Science from Karlsruhe Institute of Technology (KIT) in July 2018 under the supervision of Prof. J\"org Henkel.
Mrs. Khdr received her B. Sc in Computer Science from Aleppo University in Syria, with excellent grade and the first rank.
She is currently a research group leader at the Chair for Embedded Systems (CES) at KIT.
Her main research interests are resource management techniques that consider power, temperature and aging issues in embedded processors.
\end{IEEEbiography}

\begin{IEEEbiography}[{\includegraphics[width=1in,height=1.25in,clip,keepaspectratio]{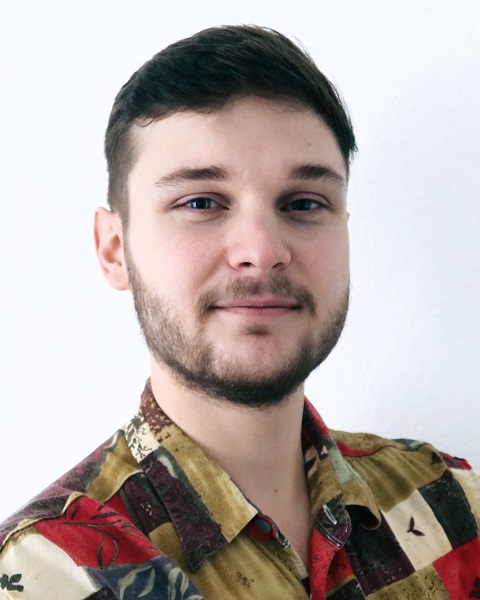}}]{Nikita Krohmer}
received a B.Sc. degree in Computer Science from Technical University of Berlin in 2018 and a M.Sc. degree in Computer Science from Karlsruhe Institute of Technology in 2021.
His research interests lie in embedded machine learning and applied artificial intelligence.
\end{IEEEbiography}

\begin{IEEEbiography}[{\includegraphics[width=1in,height=1.25in,clip,keepaspectratio]{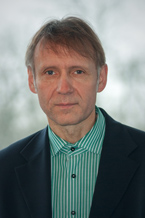}}]{J\"org Henkel}
received the Diploma and Ph.D.\ (summa cum laude) degrees from the Technical University of Braunschweig, Germany.
He was a Research Staff Member with NEC Laboratories, Princeton, NJ, and is currently the Chair Professor of embedded systems with the Karlsruhe Institute of Technology, Karlsruhe, Germany.
His research focus is on co-design for embedded hardware/software systems.
Dr.\ Henkel has received six best paper awards from major CAD conferences.
He served as the Editor-in-Chief for the ACM Transactions on Embedded Computing Systems and IEEE Design\&Test.
He has led several conferences as a General Chair incl.\ ICCAD, ESWeek etc.
He coordinates the DFG Program SPP 1500 “Dependable Embedded Systems” and is a site coordinator of the DFG TR89 collaborative research center on “Invasive Computing.”
He is the Chairman of the IEEE Computer Society, Germany Chapter.
He is a Fellow of the IEEE.
\end{IEEEbiography}



\vfill


\end{document}